# Handling Confidential Data on the Untrusted Cloud: An Agent-based Approach

Ernesto Damiani
Department of Information Technology
Università degli Studi di Milano
Milano, Italy
ernesto.damiani@unimi.it

Francesco Pagano
Department of Information Technology
Università degli Studi di Milano
Milano, Italy
francesco.pagano@unimi.it

*Abstract—* Cloud computing allows shared computer and storage facilities to be used by a multitude of clients. While cloud management is centralized, the information resides in the cloud and information sharing can be implemented via off-the-shelf techniques for multiuser databases. Users, however, are very diffident for not having full control over their sensitive data. Untrusted database-as-a-server techniques are neither readily extendable to the cloud environment nor easily understandable by non-technical users. To solve this problem, we present an approach where agents share reserved data in a secure manner by the use of simple grant and revoke permissions on shared data.

## I. INTRODUCTION

Cloud computing is the commercial evolution of grid computing [21]; it provides users with readily available, pay-as-you-go computing and storage power, allowing them to dynamically adapt their IT (Information Technology) costs to their needs. In this fashion, users need neither costly competence in IT system management or huge investments in the start-up phase in preparation for future growth.

While the cloud computing concept is drawing much interest, several obstacles remain to its widespread adoption including:
- Current limits of ICT infrastructure: availability, reliability and quality of service;
- Different paradigm of development of web applications with respect to those used for desktop applications;
- Privacy risks for confidential information residing in the cloud.

Hopefully, the first obstacle will diminish over time, thanks to the increasingly widespread availability of the network; the second will progressively disappear by training new developers and retraining the older; the third issue however, is far from being solved and may impair very seriously the real prospects of cloud computing.

In this paper, we illustrate some techniques for providing data protection and confidentiality in outsourced databases (Section II) and then we analyze some possible pitfalls of these techniques in Cloud Computing (Section III), which bring us to propose a new solution based on multi-agent systems (Section IV).

## II. THE PROBLEM OF PRIVACY

The cloud infrastructure can be accessible to public users (Public Cloud) or only to those operating within an organization (Private Cloud) [1]. Generally speaking, external access to shared data held by the cloud goes through the usual authentication authorization and communication phases. The access control problem is well-known in the database literature and available solutions guarantee a high degree of confidence.

However, the requirement that outsourced data cannot be accessed or altered by the maintainer of the datastore is not met as easily, especially on public clouds like Google App Engine for Business, Microsoft Azure Platform or Amazon EC2 platform.

Indeed, existing techniques for managing the outsourcing of data on untrusted database servers [11] [12] cannot be straightforwardly applied to public clouds, due to several reasons:
- The physical structure of the cloud is, by definition, undetectable from the outside: who is really storing the data?
- The user often has no control over data replication, i.e., how many copies exist (including backups) and how are they managed?
- The lack of information on the geographical location of data (or its variation over time) may lead to jurisdiction conflicts when different national laws apply.

In the next section, we will briefly summarize the available techniques for data protection on untrusted servers, and show how they are affected by the problems outlined above.

### A. Data Protection

To ensure data protection in outsourcing, the literature reports three main techniques [4]:
- Data encryption [13];
- Data fragmentation and encryption [14];
  - non-communicating servers [15][16];
  - unlinkable fragments [17];
- Data fragmentation with owner involvement [18].

*1) Data encryption*

To prevent unauthorized access by the datastore manager (DM) managing the outsourced RDBMS (Relational Data Base Management Systems), the data is stored encrypted.

Obviously, the encryption keys are not known to the DM and they are stored apart from the data. The RDBMS receives an encrypted database and it works on meaningless bit-streams that only the clients, who hold the decryption keys, can interpret correctly.

Note that decryption keys are generated and distributed to trusted clients by the data owner or by a trusted delegate.

Encryption can occur with different levels of granularity: field, record, table, db. For efficiency reasons, normally, the level adopted is the record (tuple in relational databases).

Of course, because the data is encrypted, the DBMS cannot index it based on plaintext and therefore it cannot resolve all queries. Available proposals tackle this problem by providing, for each (encrypted) field to be indexed, an additional indexable field, obtained by applying a non-injective transformation $f$ to plaintext values (e.g., a hashing of the field's content). This way, queries can be performed easily and with equality constraints, although with a precision $< 1$ (to prevent statistical data mining). The trusted client, after receiving the encrypted result set for the query, will decrypt and exclude spurious tuples. In this setting, however, it is difficult to answer range queries, since $f$ in general will not preserve the order relations of the original plaintext data. Specifically, it will be impossible for the outsourced RDBMS to answer range queries that cannot be reduced to multiple equality conditions (e.g., $1<=x<=3$ can be translated into $x=1$ or $x=2$ or $x=3$). In literature, there are several proposals for $f$, including:

    1. *Domain partitioning* [22]*:* the domain is partitioned into equivalence classes, each corresponding to a single value in the codomain of $f$;

    2. *Secure hashing* [11]: secure one-way hash function, which takes as input the clear values of an attribute and returns the corresponding index values. $f$ must be deterministic and non-injective.

To handle range queries, a solution, among others, is to use an encrypted version of a B $\pm$ tree to store plaintext values, and maintain the values order. Because the values have to be encrypted, the tree is managed at the Client side and it is read-only in the Server side.

*2) Data fragmentation*

Normally, of all the outsourced data, only some columns and/or some relations are confidential, so it is possible to split the outsourced information in two parts, one for confidential and one for public data. Its aim is to minimize the computational load of encryption/decryption.

  *a) Non-communicating servers*

In this technique, two *split databases* are stored, each in a different untrusted server (called, say, $S_1$ and $S_2$). The two untrusted servers have to be independent and non-communicating, so they cannot ally themselves to reconstruct the complete information. In such situation, the information may be stored in plaintext in each server.

With this approach, each Client query need be decomposed in two subqueries: one for $S_1$ and one for $S_2$. The resulting sets have to be related and filtered, later, at Client level.

  *b) Unlinkable fragments*

In reality, it is not easy to ensure that split servers do not communicate; therefore the previous technique may be inapplicable. A possible remedy is to divide information in two or more fragments. Each fragment contains all the fields of original information, but some are in clear while the others are encrypted. To protect encrypted values from frequency attacks, a suitable *salt* is applied to each encryption. Fragments are guaranteed to be unlinkable (i.e., it is impossible to reconstruct the original relation and to determine the sensitive values and associations without the decrypting key). These fragments may be stored in one or more servers.

Each query is then decomposed in two subqueries:
- The first, on the Server, chooses a fragment (all fragments contain the entire information) and selects tuples from it according to clear values and returns a result set where some fields are encrypted;
- The second, on Client (only if encrypted fields are involved in the query), decrypts the information and removes the spurious tuples according to encrypted values.

*3) Data fragmentation with owner involvement*

Another adaptation of non-communicating servers consists of storing locally the sensitive data and relations, while outsourcing the storage of the generic data. So, each tuple is split in a server part and in a local part, with the primary key in common. The query is then resolved as shown above.

B. *Selective access*

In many scenarios, access to data is selective, with different users enjoying different views over the data. Access can discriminate between read and write of a single record or only a part of it.

An intuitive way to handle this problem is to encrypt different portions of data with different keys that are then distributed to users according to their access privileges. To minimize overhead we want that:
- No more than one key is released to each user;
- Each resource is encrypted not more than once.

To achieve these objectives, we can use a hierarchical organization of keys. Basically, users with the same access privileges are grouped and each resource is encrypted with the key associated with the set of users that can access it. In this way, a single key can be possibly used to encrypt more than one resource.

*1) Dynamic rights management*

Should the user's rights change over time (e.g., the user changes department) it is necessary to remove that user from a group/role as follows:
- Encrypt data by a new key;
- Remove the original encrypted data;
- Send the new key to the rest of the group.

Note that these operations must be performed by data owner because the untrusted DBMS has no access to the keys. This active role of the data owner goes somewhat against the reasons for choosing to outsource data in the first place.

*a) Temporal key management*

An important issue, common to many access control policies, concerns time-dependent constraints of access permissions. In many real situations, it is likely that a user may be assigned to a certain role or class for only a certain period of time. In such case, users need a different key for each time period. A time-bound hierarchical key assignment scheme is a method to assign time-dependent encryption keys and private information to each class in the hierarchy in such a way that key derivation also depends on temporal constraints. Once a time period expires, users in a class should not be able to access any subsequent keys if not authorized to do so [7].

*b) Database replica*

In [5], the authors, exploiting the never ending lower price-per-byte, propose to replicate $n$ times the source database, where $n$ is the number of different roles having access to the database. Each database replica is a view, entirely encrypted using the key created for the corresponding role. Each time that a role is created, the corresponding view is generated and encrypted with a new key expressly generated for the newly created role. Users do not own the real key, but receive a token that allows them to address a cipher demand to a set KS of key servers on the cloud.

C. *A document base sample: Crypstore*

An example of data protection implementation by data encryption is Crypstore. It is a non-transactional architecture for the distribution of confidential data. The Storage Server contains data in encrypted form, so it cannot read them. User who wants to access data is authenticated at the Key Servers with the certificate issued by the Data Administrator and requires the decryption key. The Key Servers are $N$ and, to ensure that none of them knows the whole decryption key, each of them contains only a part of the encryption key. To rebuild the key, only $M$ ($<N$) parts of key are needed; redundancy provides greater robustness to failures and attacks (e.g., Denial of Service attacks).

In practice, it is an application of the time-honored "divide and conquer" technique, where data is separated from decryption keys.

Here the privacy is not entirely guaranteed because, theoretically at least, the owner of Key Servers and the Storage Server may agree to overcome the limitations of the system. The only way to exclude the (remote) possibility is to have trusted Key Servers, but if so, it would be useless to distinguish the two structures and we could take data directly, as plaintext, to a trusted storage. Such criticism applies however only in theory because, in practice, the probability of such an agreement decreases with the number of players involved.

### III. PRIVACY WITHIN THE CLOUD

All techniques discussed above are based on data encryption and/or data fragmentation using full separation of roles and of execution environments between the user and the datastore (and possibly the keystore) used to manage the outsourced data.

Let us now compare the assumptions behind such techniques with two of the basic tenets of current cloud computing architectures: data and applications being on the "same side of the wall", and data being managed via semantic datastores rather than by a conventional RDBMS.

A. *On the same side of the wall*

*Ubiquitous access* is a major feature of cloud computing architectures. It guarantees that cloud application users will be unrestrained by their physical location (with internet access) and unrestrained by the physical device they use to access the cloud.

To satisfy the above requirements (in particular the second), we normally use thin clients, which run cloud applications remotely via a web user interface.

The three main suppliers of Public Cloud Infrastructure (Google App Engine for Business, Amazon Elastic Compute Cloud and Windows Azure Platform) all include a datastore, and an environment for remote execution summarized in Tables I and II:

TABLE I. DATASTORE SOLUTIONS USED BY PUBLIC CLOUDS

| Environment | Datastore |
|---|---|
| Google | Bigtable |
| Amazon | IBM DB2 |
|  | IBM Informix Dynamic Server |
|  | Microsoft SQLServer Standard 2005 |
|  | MySQL Enterprise |
|  | Oracle Database 11g |
|  | Others installed by users |
| Microsoft | Microsoft SQL Azure |

TABLE II. EXECUTION ENVIRONMENTS USED BY PUBLIC CLOUDS

| Environment | Execution environment |
|---|---|
| Google | J2EE (Tomcat + GWT) |
|  | Python |
| Amazon | J2EE (IBM WAS, Oracle WebLogic Server) and others installed by users |
| Microsoft | .Net |

In all practical scenarios, public cloud suppliers handle both data and application management.

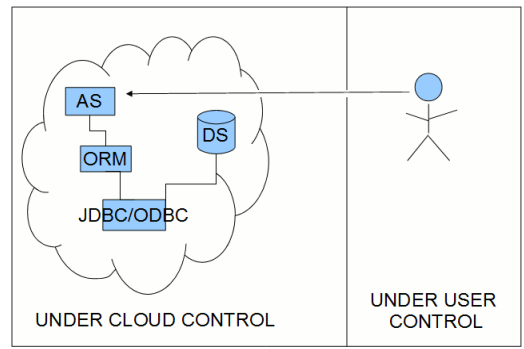

Figure 1. The wall

If the cloud supplier is untrustworthy, she can intercept communications, modify executable software components (e.g., using aspect programming), monitor the user application memory, etc.

Hence, available techniques for safely outsourcing data to untrusted DBMS no longer guarantees the confidentiality of data outsourced to the cloud.

The essential point consists in having the data and the user interface application logic *on the same side of the wall*. This is a major difference w.r.t. the outsourced database scenarios, where presentation was handled by trusted clients. In the end, the data must be presented to the user in an intelligible and clear form; that is the moment when a malicious agent operating in the cloud has more opportunities to intercept the data. To prevent unwanted access to the data at presentation time, it would be appropriate moving the presentation logics off the cloud to a trusted environment that may be an intranet or, at the bottom level, a personal computer.

However, separating data (which would stay in the cloud) from the presentation logics may enable the creation of local copies of data, and lead to an inefficient cooperation between the two parts.

*B. Semantic datastore*

Cloud computing solutions largely rely on semantic (non-relational) DBMS. These systems do not store data in tabular format, but following the natural structure of objects. After more than twenty years of experimentation (see, for instance, [8] for the Galileo system developed at the University of Pisa), today, the lower performance of these systems is no longer a problem. In the field of cloud computing, there is a particular attention to Google Bigtable.

"*Bigtable is a distributed storage system for managing structured data that is designed to scale to a very large size: petabytes of data across thousands of commodity servers. In many ways, Bigtable resembles a database: it shares many implementation strategies with databases.*" [9]

With a semantic datastore like Bigtable, there is a more strict integration between in-memory data and stored-data; they are almost indistinguishable from programmer viewpoint. There are not distinct phases when the program loads data from disk into main memory or, in the opposite direction, when program serialize data on disk. Applications do not even know where data is stored, as it is scattered over the cloud.

In such a situation, the data outsourcing techniques discussed before cannot be applied directly, because they were designed for untrusted RDBMS.

## IV. OUR APPROACH

We are now ready to discuss our new approach to the problem of cloud data privacy. We build over the notion introduced in [5] of defining a view for every user group/role, but we prevent performance degradation by keeping all data views in the user environment.

Specifically, we atomize the couple application/database, providing a copy per user. Every instance runs locally, and maintains only authorized data that is replicated and synchronized among all authorized users.

In the following subsections we will analyze our solution in detail.

*A. Information sharing by multi-agent system*

We will consider a system composed of:
1. Local agents distributed at client side;
2. A central synchronization point.

*1) The model*

In the following, we will use the term *dossier* to indicate a set of correlated information. Our data model may be informally represented by the diagram in Figure 4.

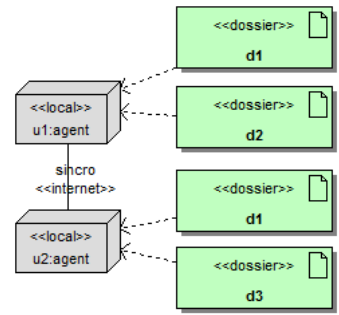

Figure 2. The model

In the model, each node represents a local, single-user application/database dedicated to an individual user ($u_n$). The node stores only the dossiers that $u_n$ owns. Shared dossiers (in this example, $d_1$) are replicated on each node. When a node modifies a shared dossier, it must synchronize, also using heuristics and learning algorithms, with the other nodes that hold a copy of it. Below we give a simple SWOT analysis of this idea.

*2) Strength/Opportunities*
- Unrestrained individual nodes, that can also work offline (with deferred synchronization);
- Simplicity of data management (single user);
- Completeness of local information.

To understand the last point, suppose that the user $u_n$ wants to know the number of the dossier she is treating. In a classic intranet solution, where dossiers would reside on their owners' servers, in addition to its database, $u_n$ should examine the data stores of all other collaborating users. With our solution, instead, $u_n$ can simply perform a local query because the dossiers are replicated at each client.

*3) Weaknesses/Threats*
- Complexity of deferred synchronization schemes [19];
- Necessity to implement a mechanism for grant/revoke and access control permissions.

This last point is particularly important and it deserves further discussion:
- As each user (except the data owner) may have partial access to a dossier, each node contains only the allowed portion of the information;

- Authorization, i.e., granting to a user $u_j$ access to a dossier $d_k$, can be achieved by the data owner simply by transmitting to the corresponding node only the data it is allowed to access;
- The inverse operation will be made in the case of a (partial or complete) revocation of access rights. An obvious difficulty lies in ensuring that data, once revoked, is no longer available to the revoked node. This is indeed a moot point, as it is impossible – whatever the approach - to prevent trusted users from creating local copies of data while they are authorized and use them after revocation.

B. *Proposed solution*

We are now ready to analyze in detail our solution. To simplify the discussion, we introduce the following assumptions:
- Each dossier has only one owner;
- Only the dossier's owner can change it.

Those assumptions allow the use of an elementary cascade synchronization in which the owner will submit the changes to the receivers.

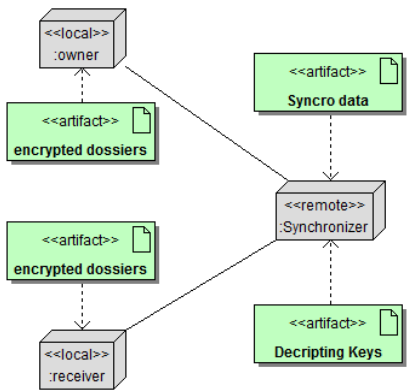

Figure 3.  Deployment diagram of multi-agent system

Our solution consists of two parts: a trusted client agent and a remote untrusted synchronizer.

The client maintains local data storage where:
- The dossiers whom he owns are (or at least can be) stored as plaintext;
- The others, instead, are encrypted, each with a different key.

The Synchronizer stores the keys to decrypt the shared dossiers owned by the local client and the modified dossiers to synchronize.

When another client needs to decrypt a dossier, he must connect to the Synchronizer and obtain the corresponding decryption key.

The data and the keys are stored in two separate entities and therefore none can access information without the collaboration of the other part.

1) *Structure*

From the architectural point of view, we divide our components into two packages, a local (client agent), which contains the dossier and additional information such as access lists, and a remote (global synchronizer), which contains the list of dossiers to synchronize, their decryption keys and the public keys of clients.

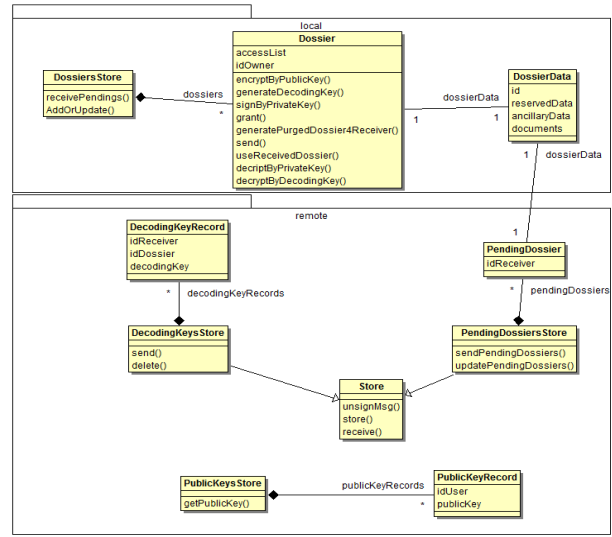

Figure 4.  Class view

2) *Grant*

An owner willing to grant rights on a dossier must follow the following sequence:

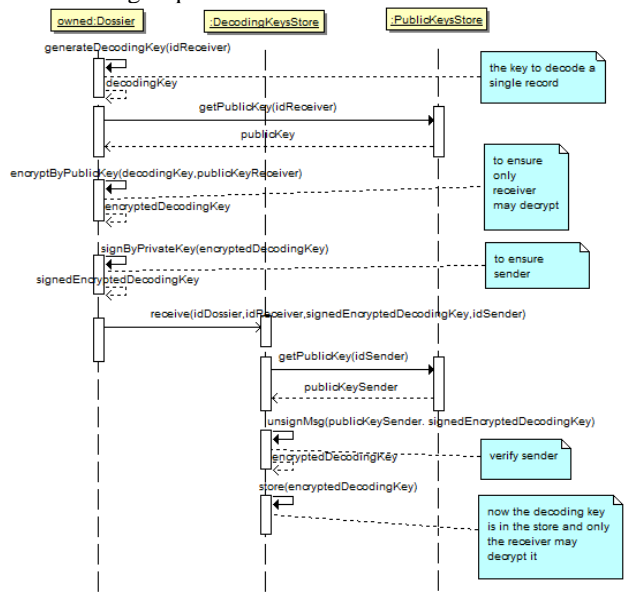

Figure 5.  Grant sequence

Namely, for each receiver, the owner:
- generates the decryption key
- encrypts it with the public key of the receiver to ensure that others cannot read it
- signs it with its private key to ensure its origin
- sends it to the Synchronizer, which verifies the origin and adds it to the storage of the decoding

keys. The key is still encrypted with the public key of the receiver, so only the receiver can read it.

*3) Send*

When an owner modifies a dossier, she sends it to the Synchronizer following this sequence:

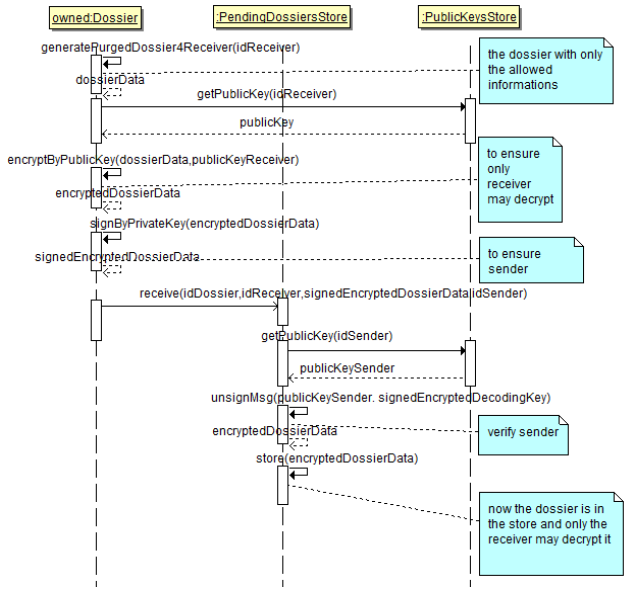

Figure 6. Send sequence

For each receiver, the owner:
- generates a "pending dossier" by removing information that the receiver should not have access to;
- encrypts it with the public key of the receiver to ensure that others cannot read it;
- signs with his own private key to certificate its origin;
- sends it to the Synchronizer, which verifies the origin and adds it to the storage of "pending dossiers". Again, the dossier is still encrypted with the public key of the receiver, so only the receiver can read it.

*4) Receive*

Periodically, each client updates un-owned dossiers by following this sequence:

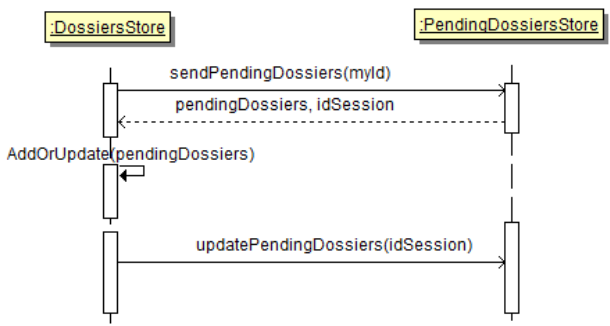

Figure 7. Receive sequence

Each client:
- requests the Synchronizer the "pending dossiers";
- modifies the local storage;
- removes from the Synchronizer the received dossiers.

*5) Use*

When a client needs to use an unowned (encrypted) dossier, the following sequence is used:

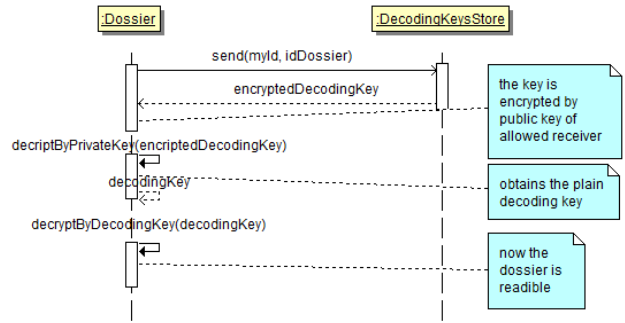

Figure 8. Use sequence

The client:
- asks the Synchronizer for the decryption key (that is encrypted by his public key);
- decrypts it with its private key;
- decrypts the dossier by the resulting decryption key.

If the decryption key does not exist, two options are available:
- the record is deleted from the local datastore because a revoke happened;
- the record remains cached (encrypted) into the local datastore because the access rights could be restored.

*6) Revoke*

To revoke access to a receiver, it is sufficient to delete the corresponding decryption key from the Synchronizer:

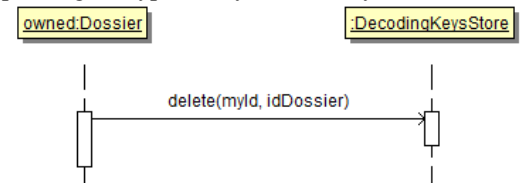

Figure 9. Revoke sequence

*7) Implementation*

We are currently implementing the proposed solution using an IMDB (in-memory database), such HyperSql (www.hsqldb.org). An in-memory database (IMDB also known as main memory database system or MMDB) is a database management system that primarily relies on main memory for computer data storage.

A HyperSql db consists of a text file containing sql instructions to:
- create structure (tables, indexes, etc.);
- populate tables.

At DBMS startup, this file is read and HyperSql creates a data model of the db into memory. At closing, the data model is serialized on the disk (actually also intermediate writes in a log file occur, to minimize the risk of data loss for sudden failure). The implementation of our solution, therefore, will consist in rewriting the load and save operations. The load function need implement the above-mentioned sequence.

*8) Future work*

In the next future, we must deepen the synchronization algorithm [23], benchmark the performance in a system under stress and use a cache of decoding time-bounded keys [6] to allow users to work offline.

## V. CONCLUSIONS AND OUTLOOK

In this paper, we discussed the applicability of outsourced DBMS solutions to the cloud and provided the outline of a simple yet complete solution for managing confidential data in public clouds.

We are fully aware that a number of problems remain to be solved. A major weakness of any data outsourcing scheme is the creation of local copies of data after it has been decrypted. If a malicious client decrypts data and then it stores the resulting plaintext data in a private location, the protection is broken, as the client will be available to access its local copy after being revoked. In [20], obfuscated web presentation logic is introduced to prevent client from harvesting data. This technique, however, exposes plaintext data to cloud provider. The manager of plaintext data is always the weak link in the chain and any solution must choose whether to trust the client-side or the server-side.

Another issue concerns the degree of trustworthiness of the participants. Indeed, untrusted Synchronizer never holds plaintext data; therefore it does not introduce an additional Trusted Third Party (TTP) with respect to the solutions described at the beginning of the paper. However, we need to trust the Synchronizer to execute correctly the protocols explained in the paper. This is a determining factor that our technique shares with competing solutions and, although an interesting topic, it lies beyond the scope of this paper.

## ACKNOWLEDGMENT

We would like to thanks Sabrina De Capitani di Vimercati and Pierangela Samarati for providing us with their seminal paper [4].

This work was partly founded by the European Commission under the project SecureSCM (contract n. FP7-213531).